\documentclass[12pt]{iopart}

\usepackage{cite,iopams}
\usepackage{latexsym}
\usepackage[dvips]{graphicx}
\usepackage{hyperref}

\newcommand{\ket}[1]{\left\vert#1\right\rangle}
\newcommand{\bra}[1]{\left\langle#1\right\vert}

\newcommand{\one}{\mbox{$1 \hspace{-1.0mm}  {\bf l}$ }} 
\newcommand{\vet}[1]{\underline{#1}}

\begin{document}

\title{A scheme for entanglement extraction from a solid}

\author{G De Chiara$^1$, {\v C} Brukner$^2$, R Fazio$^1$, G M Palma$^3$, V Vedral$^{2,4}$}
\address{$^1$ NEST-CNR-INFM \& Scuola Normale Superiore, Piazza dei
	Cavalieri 7, I-56126 Pisa, Italy}
\address{$^2$ Institut f\"ur Experimentalphysik, Universit\"at Wien, Boltzmanngasse 5, A-1090 Wien, Austria}
\address{$^3$ NEST- INFM \& Dipartimento di Tecnologie dell'Informazione, Universita' degli studi di Milano\\ via Bramante 65,
I-26013 Crema(CR), Italy}
\address{$^4$ The School of Physics and Astronomy, University of Leeds, Leeds, LS2 9JT, United Kingdom}
\ead{dechiara@sns.it}

\begin{abstract}
 Some thermodynamical properties of solids, such as heat capacity and magnetic susceptibility, have recently been shown to be
 linked to the amount of entanglement in a solid. However this entanglement may appear a mere mathematical artifact of the typical
 symmetrization procedure of many-body wave function in solid state physics.
 Here we show that this entanglement is physical demonstrating the principles of its extraction from a typical solid state system
 by scattering two particles off the system. Moreover we show how to simulate this process using present-day optical lattices technology.
 This demonstrates not only that entanglement exists in solids but also that it can be used for quantum information
 processing or for test of Bell's inequalities.
 \end{abstract}
\pacs{03.67.Mn, 03.67.-a, 03.65.Ud} 

Entanglement \cite{schroedinger} is the property of quantum systems to show correlations that cannot be explained by classical
mechanics \cite{bell}. In the last two decades entanglement has been recognized as a resource in quantum information processing,
like quantum cryptography \cite{crypto} and quantum computation, \cite{shor94,nielsenchuang}. Furthermore in recent years
the role that entanglement plays in the properties of condensed matter systems has received an increasing attention. For instance
in a typical magnetic solid the electron orbitals of different atoms overlap giving rise to an exchange interaction between the
electron spins which can be described by a Heisenberg Hamiltonian, whose thermal states are entangled \cite{bose}.

It is a common belief that entanglement cannot exist on a macroscopic scale since decoherence effects would destroy all quantum
correlations. However it has been predicted that macroscopic entanglement not only is related to critical phenomena \cite{fazio}
but also that it can exist in solids in the thermodynamic limit \cite{ghosh} even at high temperatures \cite{vlatko-highTc}.
Indeed there are many works inferring the existence of macroscopic entanglement at various temperatures up to room temperature
\cite{vedral1,vedral2,vertesi}. The evidence for this entanglement comes from experiments measuring different thermodynamic
properties such as magnetic susceptibility and heat capacity. The main aim of this article is to show that such entanglement is not an
artificial mathematical property but that it can be extracted and therefore used for quantum information processing in the same
way as heat can be extracted and used for work in thermodynamics. In particular we will show how such entanglement can be
transferred to a pair of particles and subsequently used, in principle, for quantum communication or to test the violation Bell's
inequalities. Furthermore we propose a scheme to demonstrate entanglement extraction with present-day technology using optical
lattices. In the next two paragraphs we outline the basic idea behind our proposal which is then elaborated in the remaining part
of the article.

\begin{figure}[ht]
\begin{center}
\includegraphics[scale=1]{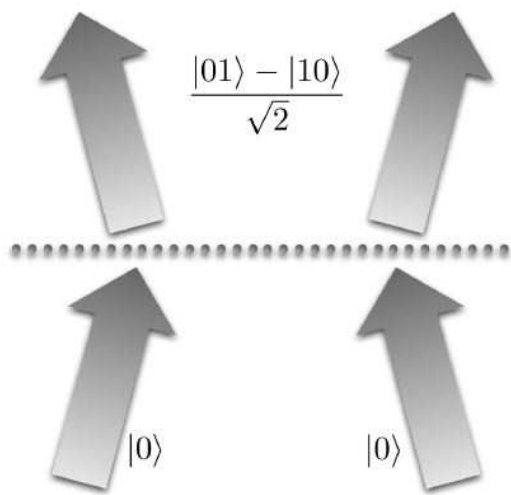}
\caption{Schematic drawing of the entanglement extraction process. Two probe particles initially prepared in a product state are
sent toward a solid whose spins are entangled. Each probe interacts with only one spin in the solid. During the collision, due to
the exchange interaction between probe and spin, their state is swapped. The two probes come out from the solid in an entangled
state.} 
\label{fig:1}
\end{center}
\end{figure}

In the ideal scenario the entanglement between a pair of spins inside a solid can clearly be transferred to another pair of probe
particles using only local swap operations. To this end one sends simultaneously a pair of probe particles toward the entangled
spin chain in such a way that each probe interacts with a different spin (cf. Fig.\ref{fig:1}). We emphasize that the two probes
do not interact with each other nor do they experience an interaction mediated by the solid (cf. \cite{braun}). The entanglement
between the probes has been extracted from the spin chain and cannot exist without entanglement in the chain. This is a genuine
non local process between the two probes like in the case of entanglement swapping \cite{entswap}. We mention that in the
continuum limit our procedure can be used  to extract entanglement from vacuum (some steps in this direction have been taken by
\cite{reznik}). In practice, however, the spin chain is in a mixed entangled state and the scattering interaction between probes
and spins in the chain can only partially swap their state. Is entanglement extraction still possible under such realistic
conditions? In this article we answer affirmatively to this question. We will consider only common physical interactions like the
Heisenberg or the XY ones. We will show that in general, after one scattering event, entanglement swapping is only partial. In
this case it is interesting to investigate whether by repeating this collision procedure with the same probes, one can achieve a
better entanglement extraction.

\begin{figure}[ht]
\begin{center}
\includegraphics[scale=1]{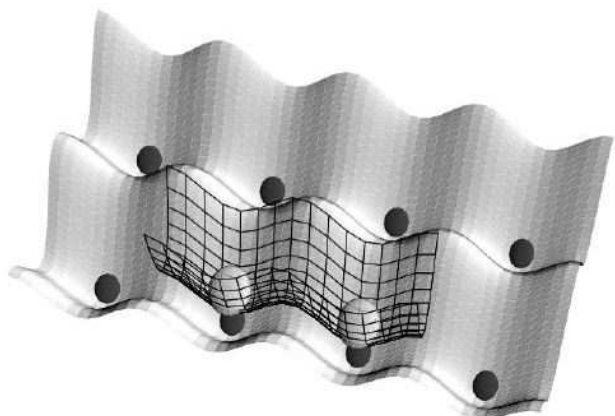}
\vspace{-0.5cm} \caption{Simulation of the entanglement extraction using optical lattices. Cold neutral atoms (black) are loaded
onto an optical lattice generated by counter-propagating lasers in such a way that each potential well is occupied by only one
atom. These atoms represent the spins degrees of freedom in a typical solid. To the first potential a second one is superimposed
(wireframe) onto which different atoms (white), called markers, are loaded. It is possible to move the markers across the solid by
moving adiabatically the second potential. This can be efficiently done by varying laser parameters of the second lattice. A
collision between a marker and a spin can be produced by moving the marker into the same potential well of the spin.}
\label{fig:lattice}
\end{center}
\end{figure}

The most natural way to extract entanglement from entangled electron spins in solids would be to scatter pairs of neutrons off the
solid. We will present a proposal for simulation of this process of entanglement extraction with optical lattices  \cite{bloch}
that we believe can be implemented at present. Optical lattices are a very useful tool for simulating many quantum effects in
solids and have been employed in demonstrating quantum phase transition  \cite{bloch}, and quantum gates \cite{contr-coll}. For
our purposes we need to simulate both the interaction in the solid system as well as the interaction with the external probes.
 Hamiltonians of entangled spin chains or ladders can be realized using cold neutral atoms trapped in potential wells
 generated by counter-propagating lasers \cite{cirac,duan,cirac1}. Bosonic atoms (black in Fig.~\ref{fig:lattice})
 are loaded onto the lattice in such a way that there is only one atom per lattice site and that the hopping between
 neighboring sites is inhibited. Long-lived atomic states are used to simulate the electronic spin in a typical solid.
 The interaction between atoms can be varied by adjusting the laser parameters and by means of electric and magnetic fields.
 In this way one can produce a variety of common spin Hamiltonians.
 As probe particles one can use different atoms (white in Fig.~\ref{fig:lattice}), called  marker qubits \cite{zoller}.
 The main idea is  to construct first the optical lattice, loading the atoms in a register that constitutes the simulated solid and
 then to superimpose a second optical lattice loading auxiliary atoms onto it as shown in Fig.~\ref{fig:lattice}.
 The auxiliary marker atoms need not to be of the same atomic species as the atoms in the solid.
 The marker atoms can be moved through the solid by varying adiabatically laser parameters of the second optical lattice.
 These operations do not affect the register atoms which remain confined in their potential wells.
In this way collisions between probes and spins in the solid can be simulated by moving a marker atom close to a register atom
letting their orbitals overlap as shown in Fig.~\ref{fig:lattice}. This in principle can  be performed with high precision and
efficiency as in the case of cold controlled collisions \cite{contr-coll}. In this way one can realize the swap operations needed
in our protocol. Such an operation can be repeated several, times moving the markers forward to collide with other register atoms.

Let us now introduce the detailed description of  the interaction between the probes and the spins in the solid. We stress again
that all aspects of the following analysis can be implemented not only with optical lattices described above but also with any
other quantum systems capable of coherent manipulation. Two probes are sent to interact for a time $\tau$ each with a single spin
of an entangled pair in a solid as shown in Fig.~\ref{fig:1}. We suppose that the interaction Hamiltonian is given by an
anisotropic $XXZ$ model between probe $i=L,R$ and spin $j=1,2$:
\begin{equation} \label{eq:ham}
H(\lambda)_{i j} = J   \left (\sigma_x^i\sigma_x^j + \sigma_y^i\sigma_y^j +\lambda\sigma_z^i\sigma_z^j\right)
\end{equation}
where $\sigma_{x,y,z}$ are the Pauli matrices, $J$ is the coupling which we assume to be constant during the collision. In the case of a time dependent $J(t)$ our analysis still applies by making the substitution: $J t \to \int_0^t J(t') dt'$. The total
Hamiltonian is therefore $H_T=H(\lambda)_{1L}+H(\lambda)_{2R}$.  This Hamiltonian comes from the same exchange mechanism that
gives rise to the interaction between spins in the solid. This Hamiltonian can be easily implemented with optical lattices and
marker qubits in a probabilistic fashion. Here we concentrate on two important limits: $\lambda=1$, which is the Heisenberg
(exchange) interaction; $\lambda=0$, which is the XY interaction. We assume that the spin chain is initially in the ground state, however our discussion can also be extended to other out-of-equilibrium states. We also assume that during the only dynamics is due to the interaction with the probes. The reduced density matrix of the entangled pair can be written in terms of spin-spin correlation functions by using translational invariance and conservation of angular momentum. The full derivation is presented in  \cite{amico} and the result is:
\begin{equation} \label{eq:werner}
\rho= \left(\begin{array}{cccc} \frac{1}{4}+g_{zz} & 0 & 0 & 0
 \\0 & \frac{1}{4}-g_{zz} & 2g_{xx} & 0
 \\0 & 2g_{xx} & \frac{1}{4}-g_{zz} & 0
 \\0 & 0 & 0 & \frac{1}{4}+g_{zz}
 \end{array}\right)
\end{equation}
where $g_{zz}=\langle \sigma_z^1\sigma_z^2\rangle/4 $ and $g_{xx}=\langle \sigma_x^1\sigma_x^2\rangle/4$ are spin-spin correlation functions and can be numerically estimated for the Heisenberg model.
The concurrence of $\rho$, assuming $\left | g_{xx}\right |\geq g_{zz}$, is:
\begin{equation}
C=\textrm{Max}[0,-\frac{1}{2}+4\left | g_{xx}\right | -2g_{zz}]
\end{equation}

Note that the results given below apply to a large class of mixed
entangled states including, for example, Werner states.
Let us first consider what happens after a single collision between the probes in a pure product state e.g. $\ket{00}$ and the
spins when $\lambda=1,0$. To illustrate the idea let us suppose that the spins are initially in a maximally entangled state
$\ket{\psi^-}=2^{-1/2}(\ket{01}-\ket{10})$. After the collision the global state of the four particle reads (apart from a global
phase factor):
\begin{equation} \label{eq:hei_swapping}
\ket{\Psi(t)}=\cos 2J\tau \ket{00}_{LR}\ket{\psi^-}_{12}-i\sin 2J\tau \ket{\psi^-}_{LR}\ket{00}_{12}.
\end{equation}
This state is a superposition of the initial state and of the swapped state.  By tracing out the two spins one obtains the reduced
density matrix $\rho_{LR}$ of the two probes. We use the concurrence \cite{wootters} to measure the entanglement extracted. For a
density matrix of two qubits $\rho$  let us define $\tilde\rho \doteq \sigma_y\otimes\sigma_y \rho^*\sigma_y\otimes\sigma_y$  and
$R=\rho\tilde\rho$. The concurrence is defined as $ C = \max \{0,\lambda_1-\lambda_2-\lambda_3-\lambda_4 \}$ where $\lambda_i$ are
the square roots of the eigenvalues of $R$ labeled in decreasing order. The concurrence for the two probes is simply $C=\sin^2
2J\tau$ and is equal to $1$ when $J\tau=\pi/4$. Thus no matter how small $\tau$ is, it is still possible to extract some
entanglement from the singlet state.

\begin{figure}[ht]
\begin{center}
\includegraphics[scale=1]{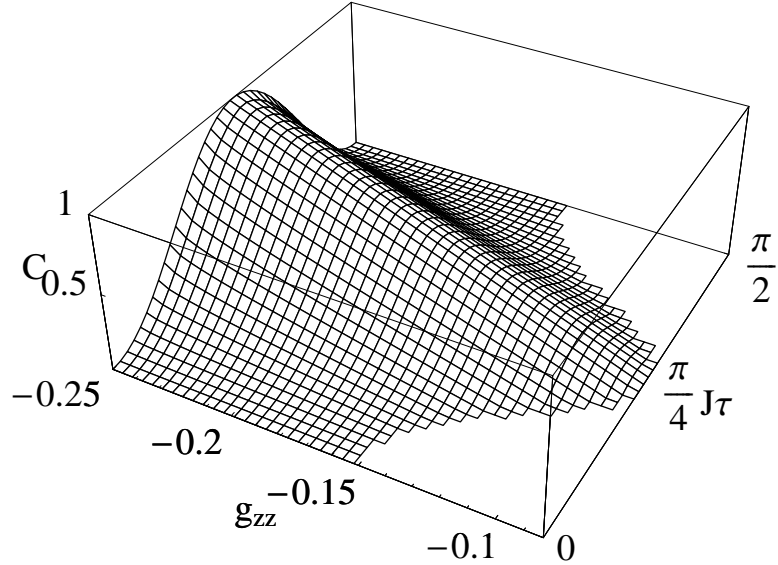}
\caption{Extracted entanglement for an Heisenberg chain, measured by the concurrence $C$ between the two probes, maximized over
the set of initial product states of the probes, after one collision as a function of $g_{zz}$ and $\tau$. Notice how $C$
increases with the correlation $g_{zz}$ and oscillates with the interaction time   $\tau$.} \label{fig:conc}
\end{center}
\end{figure}

The extracted entanglement (i.e. the concurrence between the probes), maximized over the set of initial product states of the
probes, is plotted in Fig.~\ref{fig:conc}, for $\lambda=1$, as a function of $g_{xx}=g_{zz}$, including therefore the case of
mixed states of spins in the solid. Note that $\rho$ is entangled for $-1/4\leq g_{zz}<-1/12$ . As it is shown in the plot the
entanglement between probes decays when $g_{zz}$ decreases. This is confirms the intuitive expectation that the more the spins in
the solid are entangled, the more such entanglement can be transferred to the probes.

We want to stress that for a translational invariant system the correlation functions can never reach the maximum value $g_{xx}=g_{zz}=-1/4$. Pure maximal entangled states can not exist in a homogeneous spin chain. Nevertheless in Fig.~\ref{fig:conc} we show the concurrence for all possible values of $g_{zz}$.

The state of the spins and that of the probes are swapped when $J\tau=\pi/4$ and the entanglement extracted is
maximum. In an optical lattice implementation this can be achieved in a time comparable to the trapping frequency of the atoms ($10 \mu$s for $^{87}$Rb, \cite{zoller}). In the case of solid $J$ is of the order of $1 K$ giving approximately $\tau \sim 10^{-11}$s \cite{vedral2}.
We also note that entanglement oscillates with $\tau$, a feature which reflects the fidelity of the swapping operation.
The probes are always entangled for $g_{zz}\lesssim -0.16$ no matter how small $\tau$ is. However, as shown in
Fig.~\ref{fig:conc}, for $-0.16<g_{zz}<-1/12$,  $\tau$ must be chosen appropriately in order to extract some entanglement. Notice
that obviously no entanglement can be extracted for $g_{zz}\ge -1/12$ since the two spins are not entangled. The important point
is that as long as some entanglement is shared between two spins in the solid (i.e. for $g_{zz}<-1/12$) our procedure is capable
of transferring part of it to the probes. Similar results applies also to the $XY$ model $\lambda=0$.

Once $\tau$ is fixed for a specific experimental setup, the amount of entanglement extracted is less than the entanglement in the
pair of spins as long as $J\tau<\pi/4$.
Is it possible to extract \emph{more} entanglement by repeating the collision process \emph{several} times? This can be done in
optical lattices by moving the marker qubits so they interact with different identically prepared chains (cf.
Fig.~\ref{fig:lattice}). Although this approach has some conceptual analogies with the so called homogenization process
\cite{homogen}, the two schemes are not equivalent since  the transformation induced by the interaction Hamiltonian (\ref{eq:ham})
is not a global partial SWAP transformation.
Indeed with a local interaction generated by the spin Hamiltonian ($\lambda=1$) the full transformation of the four spins is:
\begin{eqnarray}
U_{12LR}&=&PSW_{1L}\otimes PSW_{2R}, \quad \textrm{where} \label{U12LR}\\
PSW&=&e^{iJt}\left( \cos 2Jt\, \one -i\sin 2Jt\,SWAP\right), \quad \textrm{and}\\
SWAP&=&\ket{00}\bra{00}+\ket{01}\bra{10}+\ket{10}\bra{01} +\ket{11}\bra{11}
\end{eqnarray}
swaps two qubits.

\begin{figure}[ht]
\begin{center}
\includegraphics[scale=0.4]{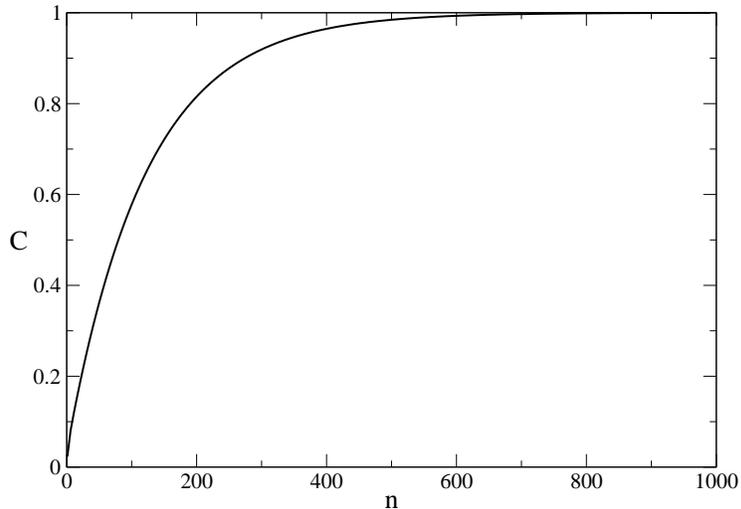}
\caption{Concurrence between the probes as a function of the number of collisions. We supposed the initial state (\ref{eq:werner}) with $\eta=0$ and $J\tau=0.2$. The data are well described by the fit function $C(n)=1-\exp(-\kappa n)$ where $\kappa=-8.3\; 10^{-3}$.}
\label{fig:4}
\end{center}
\end{figure}

It is easy to demonstrate that, for $\lambda=0$ (XY model), the state $\ket{\psi^+}_{LR}\ket{\psi^-}_{12}$ is an eigenstate of
$H_T$. It is thus a fixed point of the evolution transformation $U_{12LR}$ and the state of the probes 
approaches the state $\ket{\psi^+}$ as the number of collisions goes to infinity, independently of the input state and of how
small $\tau$ is. This is a counter-intuitive result: the state of the probes do not converge to the state of the bath but to an
orthogonal state. The reason is that $XY$ interaction does not generate a partial SWAP operation but something similar to a
partial iSWAP operation:
\begin{equation}
iSWAP=\ket{00}\bra{00}+i\ket{01}\bra{10}+i\ket{10}\bra{01} +\ket{11}\bra{11}
\end{equation}
that swaps the state of the two qubits but with a $i$ factor. The composition of two of these transformation produces a minus sign
that transform $\ket{\psi^+}$ in $\ket{\psi^-}$. In Fig.~\ref{fig:4} we show how the concurrence grows increasing the number of collision with
the spins.

For $\lambda=1$ (Heisenberg model)  there is not an eigenstate of the total Hamiltonian which is the tensor product of a 
maximally entangled state of the probes times a maximally entangled state of the two spins.

When the state of the two spins in the solid is mixed, the transformation has still a unique
fixed point \cite{homogen2,raginsky}. This follows form the Banach-Cacciopoli theorem for contractive maps. The transformation we are considering is the tensor product of two partial swap transformation which are contractive maps and thus it is itself a contractive map. The entanglement of the fixed point depends on $g_{zz}, g_{xx}$ and $\tau$. For example, for a
Heisenberg interaction, when $g_{zz}=-0.2$ and $J \tau= 0.6$, the concurrence of the fixed point is $C\simeq 0.28$. However, when the
two spins in the chain are in a mixed state, the entanglement extracted by the probes after repeated collisions is always less
than that after only one collision for $\lambda=1$ (Heisenberg interaction) and ``nearly'' always, i.e. except for mixed states
which are very close to a pure state, for $\lambda=0$ (XY interaction). Therefore repeated collisions do not increase the
extracted entanglement.

Under different hypothesis it is possible by repeated collisions to homogenize the state of the probes to the state of the spins in the solid.
By modifying the interaction between the four spins $1,2,L,R$ we can  build a Hamiltonian that generates a global SWAP transformation:
\begin{equation}
PSW_{1L,2R} = e^{iJt}\left( \cos 2Jt\, \one_{1L}\otimes\one_{2R} -i\sin 2Jt\, SWAP_{1L}\otimes SWAP_{2R}\right)
\end{equation}
This can be found by inspection and reads in the computational basis of $12LR$:
 \begin{eqnarray}
H_{12LR}&=&\ket{0000}\bra{0000}+\ket{0101}\bra{0101}+\ket{1010}\bra{1010}+\ket{1111}\bra{1111}\nonumber \\
&+&\left(\ket{0001}\bra{0100}+\ket{0010}\bra{1000}+\ket{1011}\bra{1110}+\ket{1101}\bra{0111}\right . \nonumber\\
&+&\left . \ket{0011}\bra{1100}+\ket{1001}\bra{0110}+h.c.\right )
\end{eqnarray}
Notice that the elements of the first two rows can be generated by pairwise interactions as in (\ref{eq:ham}). The elements of the third row instead are four spin processes. With this interaction it is proven by the homogeneization process \cite{homogen} that the probes state will approach the state of the spins no matter the initial state.

Until now we concentrated on extracting bipartite entanglement, however our scheme can also be adapted to extract genuine
multipartite entanglement. Let us consider $n$ spins in a chain, each scattering a spin probe. Let the initial state of the chain
be a $W$ state:
\begin{equation}\label{eq:wstate}
\ket{W_n}=\frac{1}{\sqrt n}\left(\ket{100\cdots 0}+\ket{010\cdots 0}+\cdots+\ket{000\cdots 1} \right)
\end{equation}
and the probes be in state $\ket{\vet0}=\ket{00\cdots 0}$. $W$ states can be ground states of some Hubbard related Hamiltonians
\cite{korepin}, critical spin chains \cite{fazio,chiara} and fermionic lattice models of high $T_c$ superconductivity
\cite{vlatko-highTc}. The state of the system at time $\tau$ is the analogue for $n$ spins of Eq.~(\ref{eq:hei_swapping}) with
$\ket{W_n}$ instead of $\ket{\psi^-}$. Thus the reduced density matrix for the $n$ probes reads:
\begin{equation}
\rho=\cos^2 2J\tau\ket{\vet 0}\bra{\vet 0}+\sin^2 2J\tau \ket{W_n}\bra{W_n}
\end{equation}
Notice that when $J\tau=\pi/4$ the state of the chain is fully swapped onto that of the probes as in the case with $n=2$. For any
value of $J\tau\neq k\pi/2$, where $k$ is an integer, it can be shown that the above state contains genuine multipartite
entanglement \cite{chiara} (i.e. it cannot be written as a mixture of biseparable state).

So far we have assumed that each probe interacts  with a single spin in the chain. This assumption is, in some cases, unrealistic.
For instance it is difficult, with present day technology, to address a single spin in a solid. For example, in neutron
scattering experiments, the spin of the incoming neutron interacts with the total angular momentum of a bunch of spins in the
solid \cite{rauch}. It is thus more realistic to analyze a  model in which two probes interact with several spins at once. Let us
consider a chain of $L$ spins and let us assume that each neutron, being a wavepacket of a certain width,  interacts with a
different subset of $N$ spins. We will assume that the spin of the probe is coupled to the total angular momentum of the $N$
spins, which is equivalent to assume that each probe is equally coupled to each of the $N$ spins in the chain. This model is
equivalent to the ones considered in others contexts \cite{spinstar}, and for $\lambda=0,1$ its eigenstates are known. Let us
consider a $W_L$ state as the initial state of the chain. Though this is not the most general entangled state of a spin chain, this
model can be solved analytically. The concurrence for the two probes for $\lambda=0$ is $C=\frac{2N}{L}\sin^2 2J\sqrt  N \tau$ and reaches the
maximum value $\frac{2N}{L}$ for $J\tau=\pi/(4\sqrt N)$ (the entanglement for the Heisenberg model $\lambda=1$ is greater than
zero but is always less than that of the XY model). 
 Notice that $2N/L$ is just the fraction of spins interacting with the probes and thus it is the maximum bipartite entanglement that can be extracted.
We note that these results have been obtained for a $W$ state of the
chain.  This state is symmetric under permutations of any number of spins which means that the entanglement is equally distributed among all spins. However for Heisenberg and XY chain models long range correlations are very
small and the entanglement extracted should decrease with $N$ and with the distance between the two beams of neutrons. From the above analysis it emerges that a necessary condition to extract entanglement within this realistic model is that segments of the spin chains should be entangled. This analysis can be extended  the case of non-equilibrium states and it would be interesting to see if the entanglement extracted could be larger than in the case of the $W$ state.

We have described a scheme to extract entanglement from spins in a solid by scattering probe particles. In this way macroscopic
thermal entanglement of solids can be converted into a useful resource for quantum information processing. To illustrate the
scheme we have proposed an optical lattice implementation where all aspects of our procedure can be realized with present-day
technology. In this context optical lattices offer only a simulation of entanglement extraction. This is why we have discussed how
neutron scattering can be used to achieve entanglement extraction from a real solid as well as the physical limitations of this
process. We hope that our ideas in the long run will lead to an entangling procedure for neutrons in the same way as parametric
down conversion is for creating entangled photons.


\end{document}